\documentstyle[aas2pp4]{article}
\def \as   {$^{\prime\prime}$}  
\def \cmsq           {\hbox{cm$^{-2}$}}
\def \deg          {\ifmmode ^{\circ}\else $^\circ$\fi}  
\def \etal         {{\it et~al.} }
\def \Hb           {\hbox{H$\beta$}}
\def \kms          {\rm{\hbox{km s$^{-1}$}}}
\def \Lya          {\hbox{Ly$\alpha$}}
\def \Lyb          {\hbox{Ly$\beta$}}
\def \pcc           {\hbox{cm$^{-3}$}}
\def \zaz          {{$z_a\kern -1.5pt \approx\kern -1.5pt z_e$}}
\def \zllz         {{$z_a\kern -3pt \ll\kern -3pt z_e$}}

\def \zgz          {{\kiA z\lower 3pt \hbox{a} $>$ z\lower 3pt \hbox{e}\ }}
\slugcomment{\small In press with The Astrophysical Journal}

\lefthead{Hamann et al.}
\righthead{\zaz\ Absorption Lines }

\begin{document}

\renewcommand{\baselinestretch}{1.5}

\title{\Large\bf The Nature and Origin of \zaz\ Absorption Lines \\ 
in the Redshift 0.20 Quasar, PKS 2135$-$147}

\renewcommand{\baselinestretch}{1}
\smallskip

\author{FRED HAMANN\altaffilmark{1}, E. A. BEAVER, ROSS D. COHEN, \\ 
VESA JUNKKARINEN, R. W. LYONS \& E. M. BURBIDGE}
\smallskip
\affil{The Center for Astrophysics and Space Sciences, University of 
California -- San Diego, \\ La Jolla, CA 92093-0424}

\altaffiltext{1}{fhamann@ucsd.edu} 

\begin{abstract}
\normalsize

We use new UV and optical spectra and an archival {\it HST}-WFPC2 
image to study the \zaz\ absorber in the 
$z_e\approx 0.20$ QSO PKS~2135$-$147. The UV spectra,  
obtained with {\it HST}-FOS, 
show strong \zaz\ absorption lines of \ion{C}{4}, \ion{N}{5}, 
\ion{O}{6}, \Lya\ and \Lyb . The \zaz\ line profiles are resolved, 
with deconvolved FWHM of 270 to 450~\kms . 
The Lyman decrement and the \ion{O}{6} and \ion{N}{5} doublet ratios 
indicate that there are also narrower, 
optically thick line components, and there is evidence in the 
\ion{C}{4} and \Lya\ profiles for two blended components. 
Lower limits on the total column densities are of order 
$10^{15}$ \cmsq\ for all ions. The $\sim$2:1 ratio of the \ion{C}{4} 
doublets suggests that the total \ion{C}{4} column density 
is near the lower limit. If the absorber is photoionized by the 
QSO and the derived relative columns 
in \ion{C}{4} and \ion{H}{1} are roughly correct, then the 
metallicity must be at least solar. 

The location of the \zaz\ absorber remains uncertain. 
The line redshifts indicate that the clouds have little radial motion 
(less than $\pm$200~\kms ) with respect to the QSO. This  
small velocity shift could mean that the absorber is 
outside of the deep gravitational potential of the QSO and the 
host-galaxy nucleus. Two $\sim$$L_*$ galaxies in a small 
cluster centered on PKS~2135$-$147 lie within 36$h^{-1}$ kpc 
projected distance and have redshifts consistent 
with causing or contributing to the \zaz\ lines. 
The extensive halo of the QSO's host galaxy 
could also contribute. Calculations show that the QSO is bright enough 
to photoionize gas up to \ion{O}{6} in the low-density halos of the 
host and nearby cluster galaxies. Nonetheless, there is indirect evidence 
for absorption much nearer the QSO, namely (1) the derived 
high (albeit uncertain) metallicity, (2) the relatively strong 
\ion{N}{5} absorption lines, which might be caused by a higher nitrogen 
abundance in the metal-rich gas, and (3) strong, lobe-dominated 
steep-spectrum radio emission, which is known to correlate with a much 
higher incidence of (probably intrinsic) \zaz\ lines. 

We propose that the \ion{C}{4}/\ion{N}{5}/\ion{O}{6} line ratios can be 
used as a general diagnostic of intrinsic versus intervening absorption,
as long as the line saturation effects are understood. 

\end{abstract}


%

\section{Introduction}

The working definition of associated (\zaz ) absorption lines in QSO 
spectra is that they appear 
within several thousand~\kms\ of the emission-line redshift and have 
profile widths of less than a few hundred~\kms\ 
(\cite{wey79,fol86,and87,fol88}). 
These systems are known statistically to comprise an independent 
class of absorber (\cite{fol86}), but their nature 
is generally uncertain. The relatively narrow line profiles distinguish 
\zaz\ features from the well-known broad absorption lines (BALs) that 
appear in 10\% to 15\% of QSO spectra (\cite{wey91}). However, 
there are also transitional absorbers, with intermediate line widths 
and sometimes high BAL-like velocities, that reveal a  
wide range of absorption properties between the extremes 
represented by BALs and the classic \zaz\ systems (\cite{bar97}, and 
\cite{ham97a}, 1997b and 1997c).  
BALs and transitional ``mini-BALs'' clearly form in QSO outflows 
with velocities that sometimes exceed 0.1c (see also Turnshek 1988). 
The classic 
\zaz\ systems can have a variety of origins (\cite{wey79}). 
A few \zaz\ absorbers are now known to be ``intrinsic'' to the QSO 
(forming near the QSO engine) based on (1) 
line strength variabilities, (2) multiplet ratios that imply partial 
line-of-sight coverage of the background emission source(s), 
or less conclusively, (3) smooth and relatively broad line profiles 
when observed at spectral resolutions close to the thermal speeds  
(\cite{ham97a,ham97b,bar97,ald97}). 
The known intrinsic \zaz\ systems also originate in QSO ejecta; 
they are blueshifted with respect to the emission lines and 
probably form within at least a few tens of pc of the central engine. 
Recent studies showing typically high metallicities in \zaz\ systems 
(\cite{mol94,pet94,ham97d}), consistent with the metallicities derived 
from other clearly intrinsic features like the BALs and the 
broad emission lines (\cite{ham93,fer96,tur96,ham97d}), suggest that 
{\it most} \zaz\ lines have an intrinsic origin. Similarly, the 
higher incidence of \zaz\ lines in radio-loud vs. radio-quiet 
QSOs, and the further correlation between \zaz\ occurrance and 
radio properties such as 
the spectral index and the core/lobe flux ratio, also argue for 
a frequent intrinsic origin (\cite{fol88,wil95}). 

However, particular
\zaz\ systems could still form far from the QSOs, for example 
in the QSO's own host galaxy (e.g. in the galactic halo) or in other 
galaxies along our line-of-sight. The finding that radio-loud quasars 
have more nearby galaxies than radio-quiet QSOs (\cite{smi90}), in 
addition to being more likely to have \zaz\ lines (\cite{fol88}), 
has fueled speculation that neighboring galaxies (e.g. in the 
same galaxy cluster) often cause 
\zaz\ absorption. Some studies have placed upper limits on the 
absorber gas densities that require distances not less that a few 
tens to a few hundred kpc from the QSO (\cite{wil75,mor86}). 
We will call all of these non-intrinsic systems ``intervening'' because 
they are probably similar (or identical) to the well-studied 
metal-line systems that appear far from the QSO redshifts (i.e. at 
\zllz ; \cite{ber88}).   
 
PKS~2135$-$147 (PHL~1657) is a radio-loud QSO with an emission-line 
redshift of $z_e\approx 0.20$ and strong \zaz\ absorption lines that 
were first measured in \Lya , 
\ion{C}{4} and \ion{N}{5} with the {\it International Ultraviolet Explorer}  
({\it IUE}; \cite{ber83}). The absorber is the lowest redshift \zaz\ system 
known among QSOs and, as such, could help us understand the 
possible relationship between the \zaz\ systems in 
high-redshift QSOs and similar features in low-redshift Seyfert 1 
galaxies. PKS~2135$-$147 is also interesting because  
the QSO lies near the center of a small cluster 
of galaxies (\cite{sto78,sto82,ber86a,fis96}) 
and is surrounded by diffuse emission due to its own host galaxy 
(\cite{hic87,ver90,hut92,dun93}). Redshifts published for the QSO 
(\cite{ber86a}), its host galaxy and several nearby galaxies (\cite{sto78,sto82}) 
are all similar to the \zaz\ redshift measured by {\it IUE}, leaving 
any (or all) of these objects as candidates for the \zaz\ absorber. 

Figure 1 shows an image of the PKS~2135$-$147 field  
obtained by J. Bahcall and collaborators (\cite{bah97}) on 1994 August 14 
using the Wide Field and Planetary Camera 2 (WFPC2) on board 
the {\it Hubble Space Telescope} ({\it HST}). 
The diffuse emission near the QSO is not well displayed in Figure 1 
but extends over $>$10\as\ 
diameter and thus encompasses the compact companion (labeled A in the 
figure) 1.9\as\ SE of the QSO and the nearby galaxy (labeled 1) 
5.5\as\ to the SE. The linear scale at the redshift of the QSO 
is $\sim$2.2$h^{-1}$ kpc arcsec$^{-1}$ for 
$h \equiv H_o/(100$~km s$^{-1}$~Mpc$^{-1})$. 
The luminosity surrounding PKS~2135$-$147 has a galaxy-like spectrum  
with stellar absorption features and [\ion{O}{3}] emission lines 
(\cite{sto82,hic87,sto87}). 
The compact object A lies well within the host galaxy defined by this 
diffuse luminosity and could be a secondary nucleus (\cite{sto82}). 
It has strong 
line emission and an apparently non-thermal flux distribution, 
suggesting low-level QSO-like (or Seyfert nucleus-like) activity 
(\cite{sto82}).  

\placefigure{fig1}
\placetable{tbl-1}

Table 1 lists coordinates 
for all the bright sources in Figure 1 as determined by us from the flux 
centroids. The maximum (3$\sigma$) errors in the relative positions  
should be $<$0.1\as\ unless noted otherwise in the table. The errors 
in the absolute positions could be as large as 1.5\as\ (according to 
the {\it WFPC2} Instrument Handbook, Version 3.0); however, there is 
good agreement between the quasar coordinates in Table 1 and the radio 
core position (see table legend). 
The redshift of PKS~2135$-$147 shown in Figure 1 (also Table 1) 
was measured by us from the narrow emission lines in the QSO spectrum 
(see \S2.2.1 below). The remaining redshifts are from Stockton 
(1978 and 1982) based on spectra with resolutions of 3~\AA\ for 
object A and $\sim$10~\AA\ for the galaxies 1--3. Stockton did not 
quote uncertainties, but we estimate 1$\sigma$ errors in those measurements 
of $\sim$0.5~\AA\ ($\sim$0.0001 in the redshift) for object A and 
$\sim$1.5~\AA\ ($\sim$0.0003 in the redshift) for the galaxies. 
The actual uncertainties for object A and galaxy 1 could be 
slightly larger because the line emission encompassing those 
sources is spatially blended and has a complex velocity structure 
(\cite{hic87}; see also \S3.3 below). Finally, 
we measured apparent AB magnitudes for all three of the brightest cluster 
galaxies in Figure 1 (numbers 1--3) to be $m_{AB}\approx 19.3\pm 0.1$ at 
their rest-frame wavelength of $\sim$5000~\AA . These measurements imply 
absolute B-band magnitudes of $M_B\approx -19.3\pm 0.1 + 5\log(h)$ 
and luminosities of $L\approx 0.8$~L$_*$ (for $M_* = -19.5 + 5\log[h]$ 
and no corrections for the spectral slopes; see 
Lonsdale \& Chokshi 1993). 

We obtained new {\it HST} and ground-based spectra of PKS~2135$-$147 
with the goals of determining the location of the \zaz\ 
absorber and deriving its kinematics, ionization and metal abundances. 
We will show that the {\it HST} data provide significant new information 
compared to the {\it IUE} results, but the analysis is still hindered by 
modest spectral resolutions and the resulting uncertainties in the 
redshifts, absorption line profiles and column densities. 

\section{Observations and Results}

\subsection{UV Spectroscopy with {\it HST-FOS}}

We observed PKS~2135$-$147 with the pre-COSTAR {\it HST}-Faint Object 
Spectrograph (FOS) on 1992 September 13. 
The spectra obtained through the 1\as\ aperture use 
the high resolution gratings G130H\footnote{The short-wavelength 
observation was originally scheduled for the Goddard High Resolution 
Spectrograph (GHRS) with the G140M grating, which would have 
provided higher resolution and greater sensitivity. However loss of 
operation of side 1 of GHRS necessitated the use of the 
FOS-G130H.} and G190H and provide complete 
wavelength coverage from 1140 to 2330~\AA\ at 
a resolution of $\lambda /\Delta\lambda\sim 1300$ (230~\kms ). 
(The usable spectrum ends at $\sim$1220~\AA\ where it is interrupted 
by strong geocoronal \Lya\ emission.) 
Figure 2 shows the combined {\it HST} spectrum (G130H + G190H) with the 
prominent \zaz\ absorption lines labeled. The total usable exposure times 
were 40 min with G130H and 36 min with G190H. 
The spectra were calibrated by the {\it Space Telescope 
Science Institute} using their standard ``pipeline'' procedures 
immediately after the observations. 
Those calibrations have wavelength uncertainties up to a few~\AA .
We corrected the wavelengths in the G130H spectrum by applying a 
uniform shift of $-$0.15~\AA , so that the 
Galactic absorption line \ion{Si}{2}~$\lambda$1260 is at its laboratory 
wavelength in the measured spectrum. 
This shift is consistent with the 
centroid of the much broader (and less accurately measured) geocoronal 
\Lya\ line (1215.67~\AA ). 
We corrected the G190H wavelengths by simultaneously requiring 
that (1) the one-component fits to the \zaz\ absorption lines in \ion{C}{4} 
have roughly the same redshift as \ion{N}{5} in the wavelength-corrected G130H 
spectrum (\S2.1.1 below), and (2) the narrow core of the He~II~$\lambda$1640
emission line has a redshift consistent with He~II~$\lambda$4686 and 
the other narrow emission lines in the 
Lick spectrum (\S2.2.1 below). 
These requirements both indicate a wavelength 
shift of $-$0.50~\AA . All further discussions in this paper will involve the 
wavelength-corrected spectra. 

\placefigure{fig2}

\subsubsection{Absorption Line Measurements}

Figure 3 shows the spectral regions across \Lyb -\ion{O}{6}, 
\Lya -\ion{N}{5} and \ion{C}{4} on an expanded wavelength scale, 
together with polynomial fits to the local ``continuum'' 
(which include the broad emission lines) and gaussian fits 
to the \zaz\ absorption profiles\footnote{All of the fits and 
measurements (here and in \S2.2.1 below) were performed using 
the IRAF data analysis software (provided by the National 
Optical Astronomy Observatories), including the supplemental 
program SPECFIT 
written by Gerald Kriss. Note that the use of gaussian profiles 
for fitting the unresolved or marginally resolved absorption lines is 
appropriate because the strong core of the FOS line spread 
function is well approximated by a gaussian (\cite{eva93}).}. 
Table 2 lists various line fitting results, including the centroid 
wavelengths ($\lambda_{obs}$), the absorption redshifts ($z_a$) based 
on the laboratory wavelengths in Verner \etal (1994), the 
rest equivalent widths (REW) and the full widths at half minimum 
(FWHM). We constrained the gaussian fits so that both members of the 
\ion{C}{4}, \ion{N}{5} and \ion{O}{6} doublets have the same redshift and 
FWHM. The results from fitting each line 
with a single gaussian are given in the top half of Table 2.  
\Lya\ and \ion{C}{4} have weak secondary ``notches'' in 
their blue wings and are fit better with two gaussian components. 
Attempts at fitting the 
other \zaz\ lines with two components failed to yield 
unique results. The dotted curves in Figure 3 show the two-component fits 
to \Lya\ and \ion{C}{4} and one-component fits to the other lines. 
The results from the two-component fits are listed  
in the middle of Table 2. 
The bottom of the table provides data from a one-component 
fit to the Galactic absorption line \ion{Si}{2}~$\lambda$1260, which is the 
only Galactic line detected in our spectra. 

Lines searched for but not detected in the \zaz\ system at 
REW~$<$~0.3~\AA\ (3$\sigma$) include \ion{Si}{2}~$\lambda$1260, 
\ion{C}{2}~$\lambda$1335 and \ion{Si}{4}~$\lambda\lambda$1394,1403. 
Bergeron \& Kunth (1983) reported small changes in the \Lya\ 
and \ion{C}{4} absorption lines between two {\it IUE} observations 
in 1979 and 1981. Our experience with similar {\it IUE} data suggests 
that those changes are not significant, especially if one 
allows for possible real changes in the underlying broad emission-line 
profiles. The {\it HST} measurements of the \zaz\ lines are 
the same as the overall {\it IUE} results within the uncertainties. 
We conclude that there is no convincing evidence for absorption-line 
variability. 

\placefigure{fig3}
\placetable{tbl-2}

\subsection{Optical Spectroscopy at Lick Observatory}

We obtained ground-based optical spectra of \\ PKS~2135$-$147 on  
1992 September 5 (8 days before the {\it HST} observations) 
at Lick Observatory using the Shane 3 m telescope and 
the KAST spectrograph. This spectrograph employs 
a dichroic beam-splitter to record the ``blue'' and ``red'' 
spectra simultaneously in two channels using $1200\times 400$ 
pixel Reticon CCD detectors. We used a 2\as\ entrance slit for our 
science observations and a 6\as\ slit for flux calibration 
(relative to standard stars). 
The data were reduced in the usual way using VISTA (\cite{bar93}). 
Our highest resolution observations, 
used for measuring the emission-line redshifts below, cover 
$\sim$3240--4600~\AA\ in the blue channel with a mean dispersion of 
1.13~\AA\ per pixel and a resolution of FWHM~$\sim$~2.9~\AA . 
The corresponding red-channel spectrum spans $\sim$5130--6540~\AA\ 
at 1.17~\AA\ per pixel and has resolution FWHM~$\sim$~3.0~\AA . 
These spectra are plotted in Figure 4. There is no evidence for 
\ion{Mg}{2}~$\lambda\lambda$2796,2803 absorption in the \zaz\ 
system above an upper limit of REW~$\sim$~0.4~\AA . 

\placefigure{fig4}

\subsubsection{Emission Line Measurements}

Table 3 lists the centroid wavelengths and redshifts of 
emission lines in the {\it HST} (He~II~$\lambda$1640 only) and 
Lick spectra. The emission-line spectrum of PKS~2135$-$147 has a 
mixture of distinct broad and narrow line components; these are listed 
separately in the top and bottom portions of the table. We 
measured the line centroids using only the upper $\sim$80\% of the 
profiles (i.e. the line flux that lies above the continuum 
by more than $\sim$20\% of that line's peak height). This 
procedure yields robust redshifts that are not sensitive to the 
uncertain continuum placement. [\ion{O}{3}]~$\lambda$4959 and the narrow \Hb\ 
component both sit atop a broad \Hb\ emission profile. 
We measured the redshifts in these narrow lines by treating the 
broad \Hb\ emission as part of the continuum. We measured the broad \Hb\ 
itself by interpolating a curve through the base of the narrow \Hb\ 
component and using only the upper $\sim$50\% of the broad profile to avoid 
the portions that blend with [\ion{O}{3}] in the red wing. The uncertainties 
in the wavelengths and redshifts listed in Table 3 reflect only 
the photon counting statistics. The 1$\sigma$ uncertainties in the 
wavelength calibration are an additional $\sim$0.1~\AA . 

\placetable{tbl-3}

The redshifts in Table 3 follow from the 
laboratory wavelengths in Verner \etal (1994), Osterbrock (1989), 
Wiese \etal (1966).  We adopt a mean wavelength of 3727.4~\AA\ for the 
blended [O~II] doublet, assuming the pair's intrinsic flux ratio is unity. 
The mean redshift of the narrow emission lines in Table 3, excluding 
He~II, is $z_e=0.20032\pm 0.00004$, which agrees well with the QSO 
redshifts of $0.20036\pm 0.00004$ reported by Bergeron \& Boisse 
(1986) and 0.2004 given by Stockton (1982). 

We note in passing that the ratio of [\ion{O}{3}] line fluxes,
$(f(\lambda 4959)$+$f(\lambda 5007))$/$f(\lambda 4363)$, provides 
a temperature estimate for the narrow-line region in 
PKS~2135$-$147. We measure \\
$(f(\lambda 4959)$+$f(\lambda 5007))$/$f(\lambda 4363) = 80\pm 5$ from the 
spectrum in Figure 4, which corresponds to $T_e = 14,200\pm 400$~K for 
densities $n_e\la 10^5$~\pcc\ (\cite{ost89}). This result is consistent 
with the more plentiful observations of [\ion{O}{3}] in Seyfert and 
radio galaxies 
(Tadhunter, Robinson \& Morganti 1989, \cite{sto96}), and it supports  
the general finding that the [\ion{O}{3}] temperatures are too high 
compared to simple photoionization models that assume low 
(less-than-critical) densities (e.g. \cite{wil96}).

\section{Discussion}

\subsection{Line Velocities and Redshifts}

The best one-component fits to the \zaz\ lines have redshifts 
that are larger than the QSO's forbidden-line redshift by about 150~\kms\ 
(compare Table 2 and 3). This difference is small compared to 
the absorption-line FWHMs, which range from roughly 350 to 
550~\kms . A gaussian deconvolution of these FWHMs indicates 
line-of-sight velocity dispersions of 270 to 450~\kms . 
In the best two-components fits to \Lya\ and \ion{C}{4}, the individual 
absorption features are separated by $\sim$325~\kms\ 
and straddle the QSO's forbidden-line redshift. The redshift 
separation between these absorption components also encompasses the 
redshifts of companion A and the cluster galaxies 1--3 in Figure 1. 
The redshifts alone, therefore, fail to identify the absorber. 

Nonetheless, the small velocity shift between the QSO and the \zaz\ gas 
could mean that the absorber is essentially at rest ($\pm$200~\kms ) 
with respect to the QSO and therefore outside 
the deep gravitational potential of the QSO and its host-galaxy nucleus, 
at least a kpc away. However, absorption much nearer the QSO cannot 
be ruled out. The true space velocities might be much larger if they are 
dominated by rotation or infall/outflow that is mostly perpendicular 
to our line-of-sight to the QSO. For example, in the models of 
QSO outflows driven off of accretion disks (\cite{shl85,mur95}), 
the gas velocities 
near the disk are dominated by rotation plus a small vertical component 
(perpendicular to the disk plane). Absorption lines measured along 
sightlines that graze the 
accretion disk might therefore have small line-of-sight velocities, 
even though the gas is participating in a high-velocity wind 
very near the QSO.

\subsection{Projected Distances of the Cluster Galaxies}

The projected distances of the cluster objects from PKS~2135$-$147 
are 4.0$h^{-1}$ kpc for object A and 12$h^{-1}$, 36$h^{-1}$ and 
90$h^{-1}$ kpc for galaxies 1, 2 and 3, respectively. 
Studies of low-redshift intervening (\zllz ) systems show that luminous 
($\sim$$L_*$) galaxies like those near PKS~2135$-$147 can cause 
absorption lines with REW~$\ga 1.0$~\AA\ if they are close enough 
to our sightline to the QSO. In the survey by 
Lanzetta \etal (1995), 2 of 4 galaxies at projected 
distances $\la$42$h^{-1}$~kpc but none at larger distances produced 
strong (REW~$\ga 1.0$~\AA ) \ion{C}{4} absorption. Similarly, 
5 of 5 galaxies at $\la$80$h^{-1}$~kpc and 7 of 9 at 
$\la$100$h^{-1}$~kpc gave rise to strong \Lya . For a situation 
like PKS~2135$-$147, where the QSO lies at the center of a cluster, 
we expect $\sim$50\% lower probabilities for the cluster galaxies 
producing \zaz\ lines because roughly half the galaxies will lie behind 
the QSO. Any given cluster galaxy 
has no more than a 50\% chance of producing \zaz\ lines because 
of this ambiguity. The very limited statistics provided by Lanzetta 
\etal therefore suggest that galaxies 1 and 2 near PKS~2135$-$147 each 
have a $\sim$25\% chance of contributing significantly to the \ion{C}{4} 
absorption and a $\sim$50\% chance of contributing to \Lya . 
Similarly, the more distant galaxy 3 has a $\sim$25\% chance of 
contributing to \Lya\ but essentially no chance of causing 
strong \ion{C}{4} absorption. 

These arguments based on projected distances 
suggest that the best candidate for the \zaz\ absorber is 
the diffuse halo surrounding object A and the 
QSO itself (\cite{hic87}). However, the 
likelihood of such a halo producing \zaz\ lines is 
unknown. Presumably many (most?) QSOs have host-galaxy halos 
like PKS~2135$-$147, but only $\sim$25\% of radio-loud 
and even fewer radio-quiet QSOs have strong \ion{C}{4} absorption 
lines at \zaz\ (\cite{fol88}). 

\subsection{Line Optical Depths and Column Densities}

The moderate resolution of the {\it HST}-FOS spectra ($\sim$230~\kms ) 
necessarily leads to uncertainties in the measured profiles and 
derived column densities. The full widths of the \zaz\ 
lines appear marginally resolved in these data (Table 2), 
but narrow, unresolved line components could  
``hide'' considerable column density while contributing 
little to the line equivalent widths. In particular, line components with 
velocity widths as low as  
the thermal kinetic speeds of $\sim$5 to $\sim$20~\kms\ could be 
present based on observations of intervening (\zllz ) metal-line systems 
(\cite{bla88}). The measured absorption line depths in PKS~2135$-$147, 
coupled with the \ion{N}{5} and \ion{O}{6} 
doublet ratios and the \Lya /\Lyb\ decrement, suggest that there are  
unresolved line components with large optical depths. The \ion{C}{4} 
doublet ratios are consistent with optically thin absorption, but 
that result is uncertain because the lines are blended and the shape of 
the underlying broad emission-line profile is unknown. 
The last two columns in Table 2 provide lower limits on the 
line-center optical depths ($\tau_{0}^{min}$) and ionic column densities 
($\log N^{min}$ in \cmsq ) assuming that each line is fully resolved 
with the REW and FWHM given by our gaussian fits. 
These values of $\tau_{0}^{min}$ and log~$N^{min}$ follow from a 
simple curve-of-growth analysis, where the doppler $b$-values  
relate to the fitted FWHM by $b$~=~FWHM/$\sqrt{4\ln 2}$. 
The column densities, although strictly just lower limits, 
are typical of high-redshift \zaz\ systems observed at much higher spectral 
resolutions from the ground (see \cite{ham97d} and references therein). 

\subsection{Ionization}

Bergeron \& Boisse (1986) showed that the ionizing flux from the 
QSO in PKS~2135$-$147 overwhelms the intergalactic background 
radiation field throughout the cluster environment shown in Figure 1. 
They also showed that in this environment the QSO's radiative flux can easily 
photoionize the gas up to \ion{C}{4} and \ion{N}{5} if the densities are 
typical of galactic disks or halos and the distances from the QSO are 
comparable to the projected distances in Figure 1. Our {\it HST} measurement
of strong \ion{O}{6} absorption indicates that even higher levels of 
ionization are present. The standard photoionization 
calculations plotted in Figure 2b of Hamann (1997) show that the 
ionization fractions of \ion{C}{4}, \ion{N}{5} and \ion{O}{6} 
peak at ionization parameters of $U\approx 0.01$, 0.05 and 0.25, respectively 
(where $U$ is defined as the dimensionless ratio of hydrogen-ionizing photon 
to hydrogen particle densities). Those calculations used an input 
spectrum believed to be typical of Seyfert nuclei and low-redshift QSOs 
(from \cite{mat87}). The harder incident spectrum used by Bergeron 
\& Boisse (1986) in similar calculations 
leads to slightly higher ionization states at the same 
$U$, but the main results are the same. The strong presence of 
\ion{O}{6} requires gas densities that are 5 to 10 times smaller, or 
distances from the QSO that are 2 to 3 times shorter, than those needed 
for \ion{C}{4} and \ion{N}{5} alone. 

This level of ionization appears to be common in BALs and other \zaz\ 
systems that are known to be intrinsic (\cite{wey85,kor96,ham97d}). 
However, it is also easily attained in the halos of 
the cluster galaxies near PKS~2135$-$147 via photoionization by the 
QSO. If we adopt a 
distance to PKS~2135$-$147 of 645$h^{-1}$ Mpc 
and scale the Mathews \& Ferland (1987) 
spectrum to match the observed fluxes in Figure 2, we find that the 
ionization parameter is related to the gas density, $n_H$, and 
the distance from the QSO, $D$, by, 
\ $U\approx 0.25h^{-2}{\left({{0.06{\rm cm}^{-3}}\over 
n_H}\right)}{\left({{20 {\rm kpc}}\over D}\right)}^2$. The value of 
$U\approx 0.25$ needed for strong \ion{O}{6} is therefore within 
reach at low halo-like densities in the cluster galaxies. 
The most recent observations of intervening (\zllz ) metal-line absorbers 
show that ionizations up to \ion{O}{6} are commonplace even without 
the radiative flux of a nearby QSO (\cite{lu93,bur96}). 
Therefore the detection of strong \ion{O}{6} lines 
provides no insight into the location of the absorber.  

The similar column densities derived for   
\ion{C}{4}, \ion{N}{5} and \ion{O}{6} in Table 2 suggest that the 
absorber has a range of ionization states. A lower 
limit of $U\approx 0.003$ follows from the non-detections 
of \ion{C}{2} and \ion{Mg}{2}  
(see Fig. 2b in \cite{ham97d}). The larger derived column 
densities and stronger evidence for saturation in \ion{N}{5} and 
\ion{O}{6} compared to \ion{C}{4} suggest that higher 
ionizations dominate. We can place an approximate upper 
limit on the ionization parameter by assuming (for the moment) that 
the absorber has a single ionization state. This upper 
limit follows from the minimum \ion{H}{1} column density in Table 2 
and the ($\sim$2$\sigma$) upper limit on the total hydrogen column density 
$N_H\la 10^{20}$~\cmsq\ from soft X-ray observations (Rachen, Mannheim \& 
Biermann 1996; for solar abundances). Together these results imply 
a neutral hydrogen fraction of \ion{H}{1}/H~$\ga 3\times 10^{-6}$, 
which corresponds to $U\la 1$ (again, by Fig. 2b in \cite{ham97d}). 
Of course, a multi-phase absorber could still have components 
with little \ion{H}{1} and much higher $U$. Future observations at shorter 
wavelengths could directly test for higher ionization components via the 
\ion{Ne}{8}~$\lambda\lambda$770,780 doublet, which has been measured now 
in two other \zaz\ systems (\cite{bea91,ham95,ham97c}). 

\subsection{Metal Abundances}

If we assume that at least the relative column densities in 
Table 2 are approximately correct, such that any saturation effects are 
the same for all lines, we can place lower limits on the 
metal abundances even with no knowledge of the ionization. 
Photoionization calculations (e.g. \cite{ham97d,ber86b}) show that 
the ionization corrections needed to convert the 
column density ratios (for example $N$(\ion{C}{4})/$N$(\ion{H}{1})) into 
abundance ratios (C/H) always attain minimum values at some finite $U$. 
We can therefore derive the {\it minimum} metal abundances by adopting the 
{\it minimum} ionization corrections. Hamann (1997) calculated minimum 
correction factors for a wide range of incident QSO spectral shapes. 
Those calculations assume that the absorber is optically thin to 
continuum radiation at visible through far-UV wavelengths, which is 
consistent with the column densities in Table 2 and with the vast 
majority of well-measured \zaz\ systems at high redshift (\cite{ham97d}).  
(The column densities in Table 2 would have to be too 
low by $\ga$2.5 dex for H~I [from \Lya ] and by $\ga$3.2 dex for the 
metal ions to violate this assumption; cf. photoionization 
cross-sections in \cite{ost89}.) The calculations are not sensitive 
to other (unknown) physical properties of the absorber. 
Applying the minimum correction factors from Figures 8 and 9 in 
Hamann (1997) to the column density ratio of 
log($N$(\ion{C}{4})/$N$(\ion{H}{1}))~$\approx$~0.1 from Table 2 implies a 
{\it minimum} abundance ratio of [C/H]~$\approx 0.2\pm 0.2$ 
(where [C/H]~=~$\log({\rm C/H}) - \log({\rm C/H})_{\odot}$). 
The $\sim$1$\sigma$ 
uncertainty in this lower limit follows from the uncertainty in the QSO's 
flux distribution at important far-UV energies (see \cite{ham97d} 
for discussion). The same analysis 
applied to \ion{N}{5} and \ion{O}{6} leads to somewhat lower minimum 
values of [N/H] and [O/H] compared to the [C/H] result.  
Keep in mind, however, that these results must be viewed with caution 
because of the uncertain column densities. 

\section{Summary and Conclusions}

We use spectroscopic and imaging observations obtained 
with {\it HST} and at Lick Observatory  
to study the \zaz\ absorber in the $z_e\approx 0.20$ QSO 
PKS~2135$-$147. The {\it HST}-FOS spectra exhibit 
strong \zaz\ absorption lines of \Lya , \Lyb , 
\ion{C}{4}~$\lambda\lambda$1548,1551,  \ion{N}{5}~$\lambda\lambda$1238,1242 
and \ion{O}{6}~$\lambda\lambda$1032,1037. The is no clear evidence for 
line variability between the {\it HST} and previous {\it IUE} observations. 
The full line widths are marginally resolved in the {\it HST} spectrum, 
with deconvolved FWHM of 270 to 450~\kms . However, 
the \ion{N}{5} and \ion{O}{6} doublet ratios, and the more uncertain 
\Lyb /\Lya\ decrement, indicate that there are unresolved, optically thick 
line components. There is also direct evidence in the \ion{C}{4} and \Lya\ 
profiles for two blended absorption components. Lower limits on 
the total column densities (from a curve-of-growth analysis) 
range from $\log N$(\ion{H}{1})~$\ga 14.7$ \cmsq\ and 
$\log N$(\ion{C}{4})~$\ga 14.8$ \cmsq\ to 
$\log N$(\ion{O}{6})~$\ga 15.3$ \cmsq . The \ion{C}{4} doublet 
ratio is consistent with no saturation, so the actual 
\ion{C}{4} column might be close to the lower limit. 

Any analyses based on these column densities must be considered tentative 
because of the possible line saturation. Nonetheless, the similar 
derived columns in \ion{C}{4}, \ion{N}{5} and \ion{O}{6} suggest that 
the absorber has a range of ionization states. If the gas is photoionized 
by the QSO, this range can be characterized by ionization parameters 
from $U\sim 0.01$ to at least 0.25. The upper limit is unknown, but a 
lower limit of $U\approx 0.003$ follows from the absence of 
\ion{C}{2} and \ion{Mg}{2} absorption. The larger minimum column 
density and stronger evidence for line saturation in \ion{O}{6} 
suggest that the higher ionization states dominate. The  
QSO could produce these ionization states by 
photoionization in environments ranging from near the central engine to the low-density halos of cluster galaxies tens of kpc away. Whatever the 
location or ionization state of the absorber, we show that 
the C/H abundance ratio must be at least solar if the gas is photoionized 
by the QSO and the derived {\it relative} column densities in \ion{H}{1} 
and \ion{C}{4} are roughly accurate. 

The source of the \zaz\ absorption remains uncertain.  
The redshifts spanned by the \zaz\ line profiles are consistent with 
contributions from the galactic halo surrounding the QSO and its 
compact companion, and/or from the halos 
of two $\sim$$L_*$ galaxies at projected distances $\leq$36$h^{-1}$~kpc. 
The only firm conclusion from the redshift analysis is that the \zaz\ 
absorber has a line-of-sight velocity that is small (less than $\pm$200~\kms ) 
with respect to the QSO. This small velocity shift could mean that the 
absorber lies outside the deep gravitational potential of the QSO and 
its host-galaxy nucleus, at least a kpc away. However, 
the \zaz\ absorber could be much closer to the QSO and the true space 
velocities could be much larger if the motions are mostly perpendicular 
to our line-of-sight (e.g. rotational). 
There is, in fact, indirect evidence favoring absorption near the 
QSO. In particular, 
PKS~2135$-$147 is a strong radio source with a steep radio spectrum 
and a lobe-dominated morphology (\cite{mor93}). These 
properties are known to correlate positively with both the strength 
and frequency of \zaz\ lines (\cite{fol88,wil95}, Barthel, Tytler \& 
Vestergaard 1997), indicating that the absorbers in those cases 
are physically associated with the QSO and/or its host galaxy. 
The recent detections of time-variable \zaz\ lines in two radio-loud 
QSOs (\cite{bar97,ald97}) suggest 
that the gas resides close to the QSO (\cite{ham97a}). In addition, 
the derived (tentative) metallicity in the PKS~2135$-$147 absorber 
is too high for an intervening galaxy halo (cf. \cite{tyt88,ber88}) 
but typical of known or suspected 
intrinsic systems (\cite{mol94,pet94,tri96,tur96,kor96,ham97d}). 
Furthermore, the strong 
\ion{N}{5} absorption compared to \ion{C}{4} and 
\ion{O}{6} is common among known or suspected intrinsic absorbers 
(see \cite{ham97d} and references therein) but extremely rare 
in intervening metal-line systems (excluding those with damped-\Lya ; 
\cite{wey81,har82,sar88,ber88,bah93,lu93,lan95,bur96}). 

Relatively strong \ion{N}{5} lines are probably a good indicator 
of intrinsic versus intervening absorption in general. Weymann \etal (1981) 
and Hartquist \& Snijders (1982) noted a strong observational trend
for much larger \ion{N}{5}/\ion{C}{4} line ratios in \zaz\ systems 
compared to \zllz . This dichotomy cannot be understood simply 
in terms of higher ionization in \zaz\ absorbers 
because recent studies show that \zllz\ 
systems also typically have strong \ion{O}{6} lines and 
therefore considerable high-ionization gas (\cite{lu93,bur96}). 
\ion{N}{5} is weak relative to both \ion{C}{4} and \ion{O}{6} in 
these systems. 
We propose that the lower \ion{N}{5}/\ion{O}{6} and \ion{N}{5}/\ion{C}{4} 
line ratios in intervening systems are caused by an underabundance of 
nitrogen (relative to solar ratios) in the low-metallicity galactic halos. 
Relatively higher nitrogen abundances and stronger \ion{N}{5} 
absorption lines should occur naturally in the metal-rich 
environments of intrinsic absorbers, e.g. in the inner regions of the 
massive galaxies that host QSOs (\cite{ham93,vil93,ham97d} and references 
therein). The combined \ion{N}{5}/\ion{O}{6} and \ion{N}{5}/\ion{C}{4} 
line ratios could, in fact, be used to measure or constrain the N/O and 
N/C abundances. We must be cautious, 
however, because saturation effects in the intrinsic systems (\S3.3) 
could mimic the proposed abundance differences by driving the intrinsic 
line ratios toward 
unity. Also, differences in the \ion{N}{5}/\ion{O}{6} and 
\ion{N}{5}/\ion{C}{4} line ratios between \zaz\ and \zllz\ systems 
might still derive from subtleties in the ionization rather than 
abundance effects. For example, in galaxy halos that are not near QSOs,
\ion{N}{5} might fall in a narrow ionization ``gap'' between a photoionized  
gas component that produces mainly \ion{C}{4} and the lower ions 
and a hot, collisionally ionized component that produces \ion{O}{6} 
(\cite{bur96}). \ion{N}{5} might then be stronger in halos near QSOs 
because of the increased {\it photo}-ionization. Such halo-near-QSO 
systems would not meet our definition of intrinsic, and would not 
be expected to have enhanced nitrogen abundances, even though their 
\ion{N}{5} lines are strong. 

Further observations are needed to clearly identify the \zaz\ 
absorber in PKS~2135$-$147. For example, 
higher resolution spectra of the QSO and the cluster galaxies 
would allow us to (1) measure and compare the redshifts more 
precisely, (2) determine whether the \zaz\ lines have narrow, 
multi-component profiles typical of intervening 
metal-line systems (\cite{bla88}) or smooth and relatively broad profiles 
like some known intrinsic systems (\cite{ham97a}, 1997b and 1997c), 
(3) carefully examine the multiplet ratios to test for partial 
line-of-sight coverage of the background light source 
as another indicator of intrinsic absorption (\cite{ham97a,ham97b,bar97}), 
and (4) derive accurate column densities for the 
abundance and ionization analyses. Monitoring the QSO for line 
strength changes, particularly in response to continuum changes, might 
not only confirm the intrinsic hypothesis but place quantitative 
constraints on the absorber's density, ionization and distance from the 
QSO (\cite{bar92,ham95,ham97a,ham97b}). 

\acknowledgments

We are grateful to the referee, T. Aldcroft, for suggestions that improved 
this manuscript. The {\it HST} spectra presented here 
were obtained through the Goddard High 
Resolution Spectrograph Guaranteed Time program. This research was supported 
by NASA grants NAG 5-1630, NAG 5-1858 and NAG 5-3234. 

\clearpage
 
\begin{deluxetable}{cccl}
\tablecolumns{4}
\tablecaption{Coordinates in the PKS~2135$-$147 Field\label{tbl-1}}
\tablewidth{0pt}
\tablehead{ & \multicolumn{2}{c}{------ Epoch 2000 ------}\\
\colhead{Source\tablenotemark{a}} & \colhead{R.A.}  & \colhead{ Dec.} 
& Comment\tablenotemark{b}}  
\startdata
PKS\tablenotemark{c} & 21 37 45.248 & $-$14 32 55.70 & QSO 
($z_e = 0.2003$)\nl
 A &  21 37 45.353 & $-$14 32 56.80 &   affected by QSO flux, 
($z_e = 0.2003$) \nl
 1 &  21 37 45.585 & $-$14 32 58.35 &  ($z_e = 0.1997$)\nl
 2 &  21 37 44.732 & $-$14 32 40.31 &  ($z_e = 0.2002$)\nl
 3 &  21 37 47.238 & $-$14 33 26.71 &   multiple flux peaks, 
($z_e = 0.2008$)  \nl
 4 &  21 37 42.969 & $-$14 32 38.99 &   core not well defined \nl
 5 &  21 37 43.366 & $-$14 32 29.76 &  \nl
 6 &  21 37 43.599 & $-$14 33 01.69 &  \nl
 7 &  21 37 44.328 & $-$14 32 49.75 &  \nl
 8 &  21 37 44.340 & $-$14 33 18.64 &  \nl
 9 &  21 37 45.223 & $-$14 32 38.27 &   point source \nl
10 &  21 37 45.277 & $-$14 32 34.68 &   faint \nl
11 &  21 37 45.935 & $-$14 32 32.75 &   \nl
12 &  21 37 46.167 & $-$14 33 17.40 &  \nl
13 &  21 37 46.483 & $-$14 33 05.67 &   point source\nl
14 &  21 37 46.511 & $-$14 32 42.39 &   blended with \#15 \nl
15 &  21 37 46.516 & $-$14 32 42.02 &   brighter of pair \nl
16 &  21 37 46.563 & $-$14 33 28.98 &  \nl
17 &  21 37 47.018 & $-$14 33 13.85 &  \nl
18 &  21 37 47.165 & $-$14 32 42.32 &   point source\nl
19 &  21 37 48.053 & $-$14 32 42.37 &   point source\nl
\enddata
\tablenotetext{a}{Galaxy designations 1--3 follow Stockton 
(1978).}
\tablenotetext{b}{Only obvious point sources (with diffraction spikes) 
are noted as such.}
\tablenotetext{c}{The coordinates of the radio core are 
RA: 21~37~45.1, Dec: $-$14~32~55.6 (\cite{mor93}).}
\end{deluxetable}

\clearpage

\begin{deluxetable}{rcccccc}
\tablecolumns{7}
\tablecaption{Absorption Line Results \label{tbl-2}}
\tablewidth{0pt}
\tablehead{
\colhead{ \ Line} & \colhead{$\lambda_{obs}$} & \colhead{$z_a$} & 
\colhead{REW} & \colhead{FWHM} & \colhead{$\tau_0^{min}$} & 
\colhead{log $N^{min}$}\\ 
\colhead{} & \colhead{(\AA )}  & \colhead{} & 
\colhead{(\AA )} & \colhead{(\kms )} & \colhead{} & \colhead{(\cmsq )} \\
\cline{1-7} \\
\multicolumn{7}{c}{One-Component Fits} 
} 
\startdata
\Lyb\ \ 1025.72 & $1231.37\pm 0.16$ & $0.20049\pm 0.00016$ & $1.83\pm 0.28$ & $709\pm 146$ & 1.0 & 15.5\nl
O VI 1031.93 & $1239.08\pm 0.08$ & $0.20074\pm 0.00008$ & $1.14\pm 0.14$ & $432\pm\phn 56$ & 1.0 & 15.1\nl
O VI 1037.62 & $1245.91\pm 0.08$ & \nodata & $1.17\pm 0.14$ & \nodata & 
1.0 & 15.4\nl
\Lya\ \ 1215.67 & $1459.64\pm 0.04$ & $0.20069\pm 0.00003$ & $1.71\pm 0.06$ & $510\pm\phn 21$ & 1.1 & 14.7\nl
N V \ 1238.82 & $1487.73\pm 0.05$ & $0.20093\pm 0.00004$ & $1.06\pm 0.08$ & $351\pm\phn 32$ & 0.9 & 14.8\nl
N V \ 1242.80 & $1492.51\pm 0.05$ & \nodata & $1.01\pm 0.09$ & \nodata & 
0.9 & 15.1\nl
C IV 1548.20 & $1859.33\pm 0.06$ & $0.20092\pm 0.00004$ & $2.01\pm 0.07$ & $549\pm\phn 27$ & 0.9 & 14.8\nl
C IV 1550.88 & $1862.42\pm 0.08$ & \nodata & $1.01\pm 0.06$ & \nodata & 
0.4 & 14.8\nl
\cutinhead{Two-Component Fits} 
\Lya\ \ 1215.67 & $1458.28\pm 0.10$ & $0.19957\pm 0.00008$ & $0.37\pm 0.10$ & $229\pm\phn 55$ & 0.4 & 13.9\nl
\Lya\ \ 1215.67 & $1459.83\pm 0.05$ & $0.20084\pm 0.00004$ & $1.32\pm 0.09$ & $381\pm\phn 28$ & 1.2 & 14.5\nl
C IV 1548.20 & $1857.43\pm 0.18$ & $0.19974\pm 0.00012$ & $0.57\pm 0.15$ & $307\pm\phn 83$ & 0.4 & 14.2\nl
C IV 1550.88 & $1860.51\pm 0.18$ & \nodata & $0.39\pm 0.16$ & \nodata & 
0.3 & 14.3\nl
C IV 1548.20 & $1859.50\pm 0.08$ & $0.20108\pm 0.00005$ & $1.27\pm 0.14$ & $330\pm\phn 30$ & 1.0 & 14.6\nl
C IV 1550.88 & $1862.59\pm 0.08$ & \nodata & $0.81\pm 0.09$ & \nodata & 
0.5 & 14.7\nl
\cutinhead{Galactic Lines} 
Si II 1260.42 & $1260.42\pm 0.11$ & $0.00000\pm 0.00008$ & $0.69\pm 0.16$ & $224\pm\phn 58$ & 0.9 & 13.7\nl
\enddata
\end{deluxetable}

\begin{deluxetable}{rcc}
\tablecolumns{3}
\tablecaption{Emission Line Redshifts \label{tbl-3}}
\tablewidth{0pt}
\tablehead{
\colhead{Line} & \colhead{$\lambda_{obs}$}  & \colhead{$z_e$} \\  
\cline{1-3} \\ 
\multicolumn{3}{c}{Narrow Lines} 
} 
\startdata
He II \ 1640.4 & $1969.2\pm 0.3$ & $0.2004\pm 0.0002$\nl
[Ne V] 3425.9 & $4112.0\pm 0.3$ & $0.2003\pm 0.0001$\nl
[O II] \ 3727.4 & $4474.1\pm 0.3$ & $0.2003\pm 0.0001$\nl
He II \ 4685.7 & $5624.3\pm 0.5$ & $0.2003\pm 0.0001$\nl
H$\beta$ \ \ \ \ 4861.3& $5835.0\pm 0.6$ & $0.2003\pm 0.0001$\nl
[O III] 4958.9 & $5952.4\pm 0.1$ & $0.2003\pm 0.0001$\nl
[O III] 5006.9 & $6010.0\pm 0.1$ & $0.2003\pm 0.0001$\nl
\cutinhead{Broad Lines}
  Mg II \ 2799.2 & $3363.1\pm 0.7$ & $0.2014\pm 0.0002$\nl
H$\beta$ \ \ \ \ 4861.3& $5844.2\pm 1.2$ & $0.2022\pm 0.0002$\nl
\enddata
\end{deluxetable}

\clearpage

\clearpage

\centerline{\bf Figure Captions}
\bigskip
\figcaption[pks2135_chart.eps]{PKS~2135$-$147 field measured with 
{\it HST}-WFPC2 and a broad-band filter (F606W) centered 
at roughly 6060~\AA . The vertical ``stretch'' is linear. The 
angular scale is in arcsec relative to the QSO position. (The 
linear scale at the QSO redshift is $\sim$2.2$h^{-1}$ kpc arcsec$^{-1}$.) 
Redshifts are shown in parenthesis for 
the QSO (this paper), the compact object A (from Stockton 1982) and 
the galaxies 1--3 (Stockton 1978). The coordinates of all labeled 
sources are listed in Table 1. Separate 
CCD images join along the vertical line at $-$36\as\ and along a 
horizontal line at +39\as . \label{fig1}}

\figcaption[specplot2.ps]{Combined {\it HST}-FOS spectrum of 
PKS~2135$-$147 obtained with the G130H and G190H 
gratings of the FOS. The associated absorption lines and the 
narrow He II emission line are labeled above. 
The spectra are shown after smoothing  
twice (G190H) and 3 times (G130H) with a 3-pixel wide boxcar function. 
The 1$\sigma$ error spectrum (not smoothed) is shown by the 
dashed curve at the bottom. The associated absorption lines and the 
narrow He II emission line are labeled above. $F_{\lambda}$ has 
units 10$^{-14}$ ergs s$^{-1}$ cm$^{-2}$~\AA$^{-1}$. The segment 
between roughly 1710 and 1725~\AA\ is a straight-line interpolation 
across a detector noise spike. \label{fig2}}

\figcaption[fitplot1.ps]{{\it HST}-FOS spectra of PKS~2135$-$147 centered 
near the \Lyb -\protect\ion{O}{6} (top panel), \Lya -\protect\ion{N}{5} 
(middle panel) and \protect\ion{C}{4} (bottom panel) absorption lines. 
The dashed curve at the bottom of each panel is the $1\sigma$ error spectrum. 
The dotted curves show the fits to the continuum and the absorption lines 
discussed in the text. The lines are labeled at the redshifts derived 
from the gaussian fits, with two components for \Lya\ and \protect\ion{C}{4}. 
The panel also shows the Galactic \protect\ion{Si}{2} 1260.42~\AA\ line used 
to set the wavelength scale.\label{fig3}}

\figcaption[specplot3.ps]{Red and blue channel Lick 
spectra of PKS~2135$-$147. The flux units are 
10$^{-15}$ ergs s$^{-1}$ cm$^{-2}$ Hz$^{-1}$. The flux scale at the 
right corresponds to the upper spectrum in the bottom panel. 
The dashed curve at the bottom of each panel is the $1\sigma$ error spectrum. The emission lines are labeled at $z_e=0.2003$. \label{fig4}}



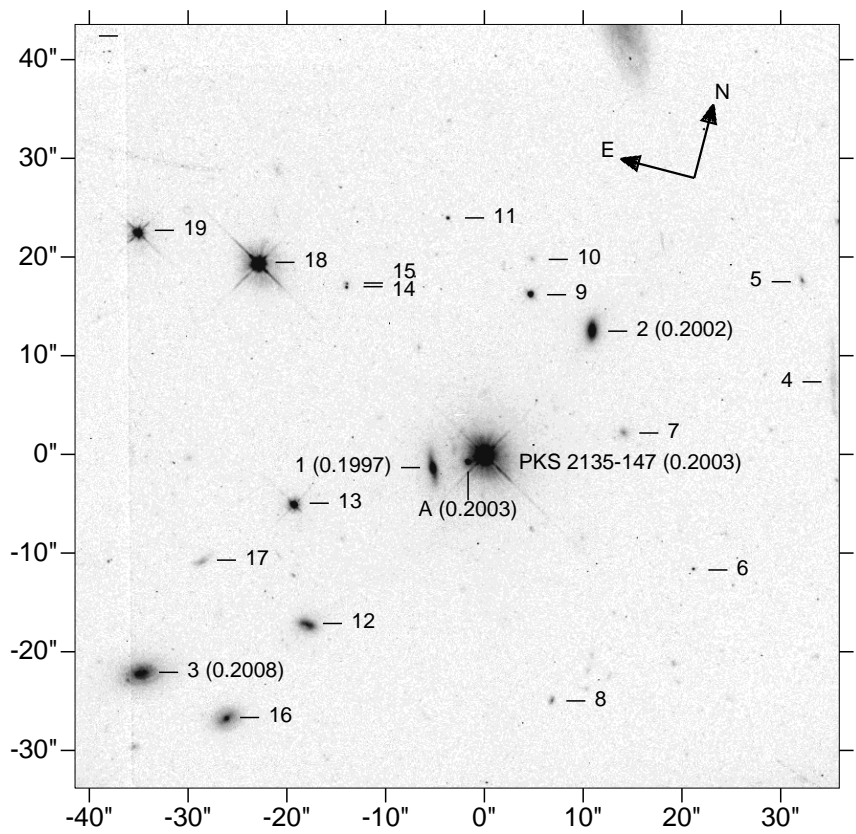
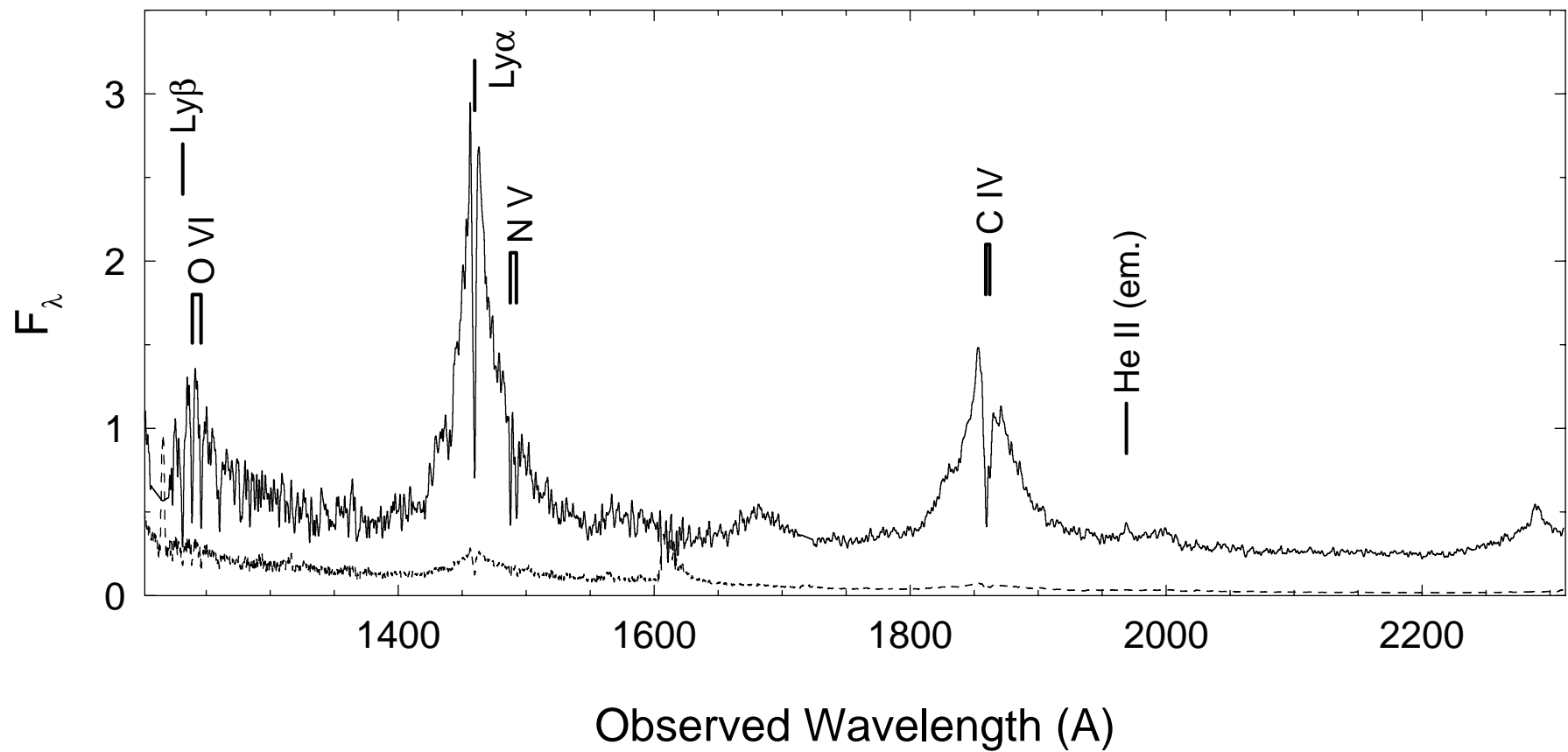
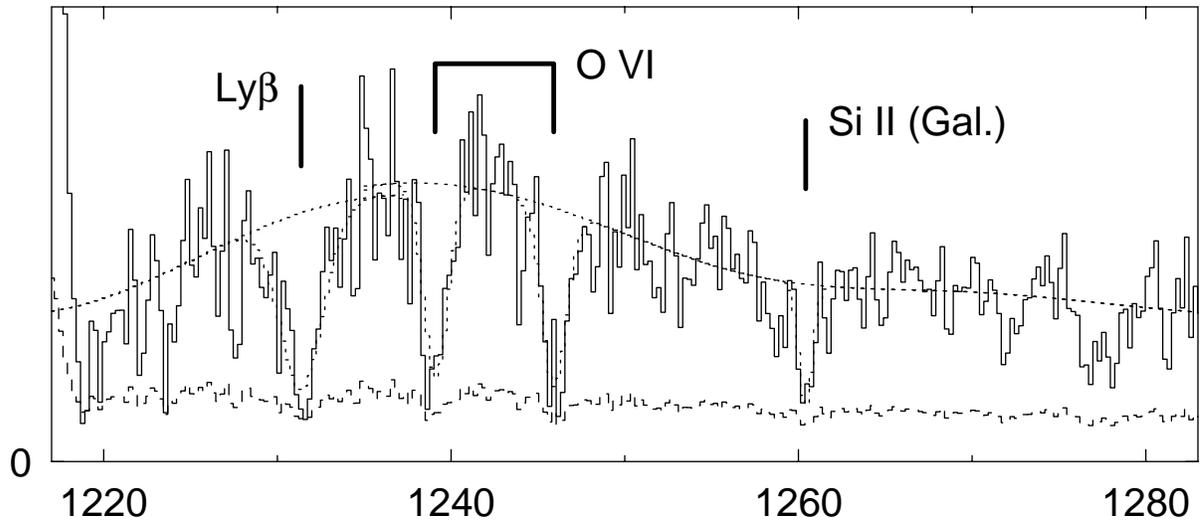
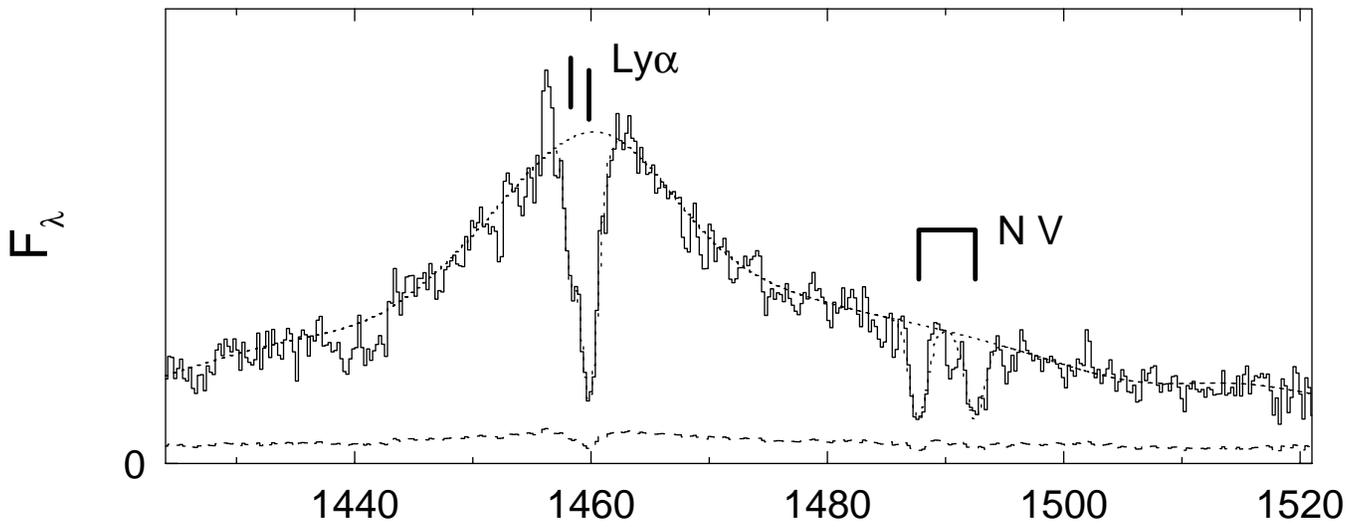
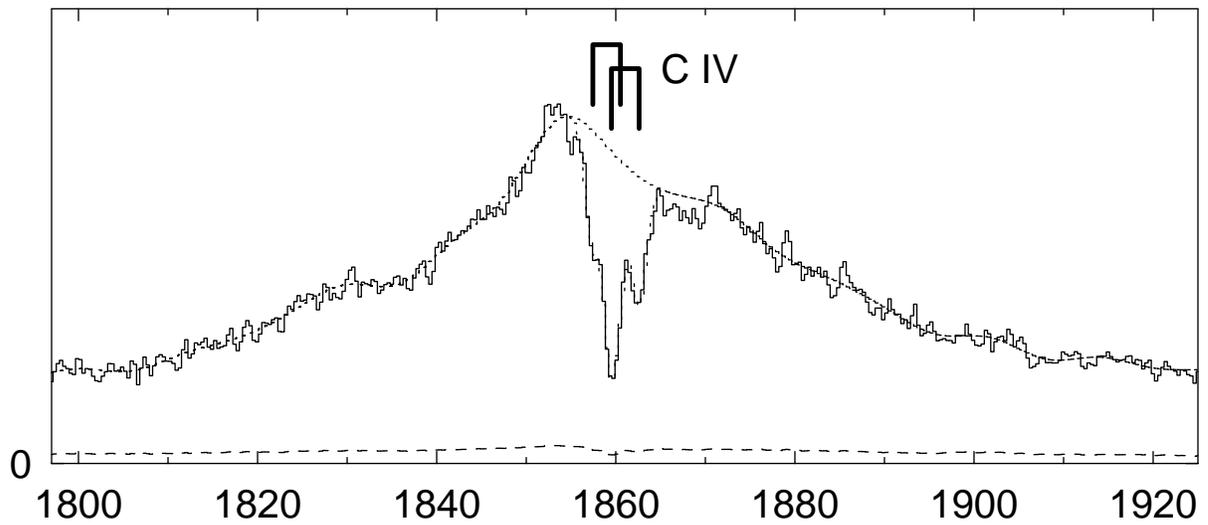
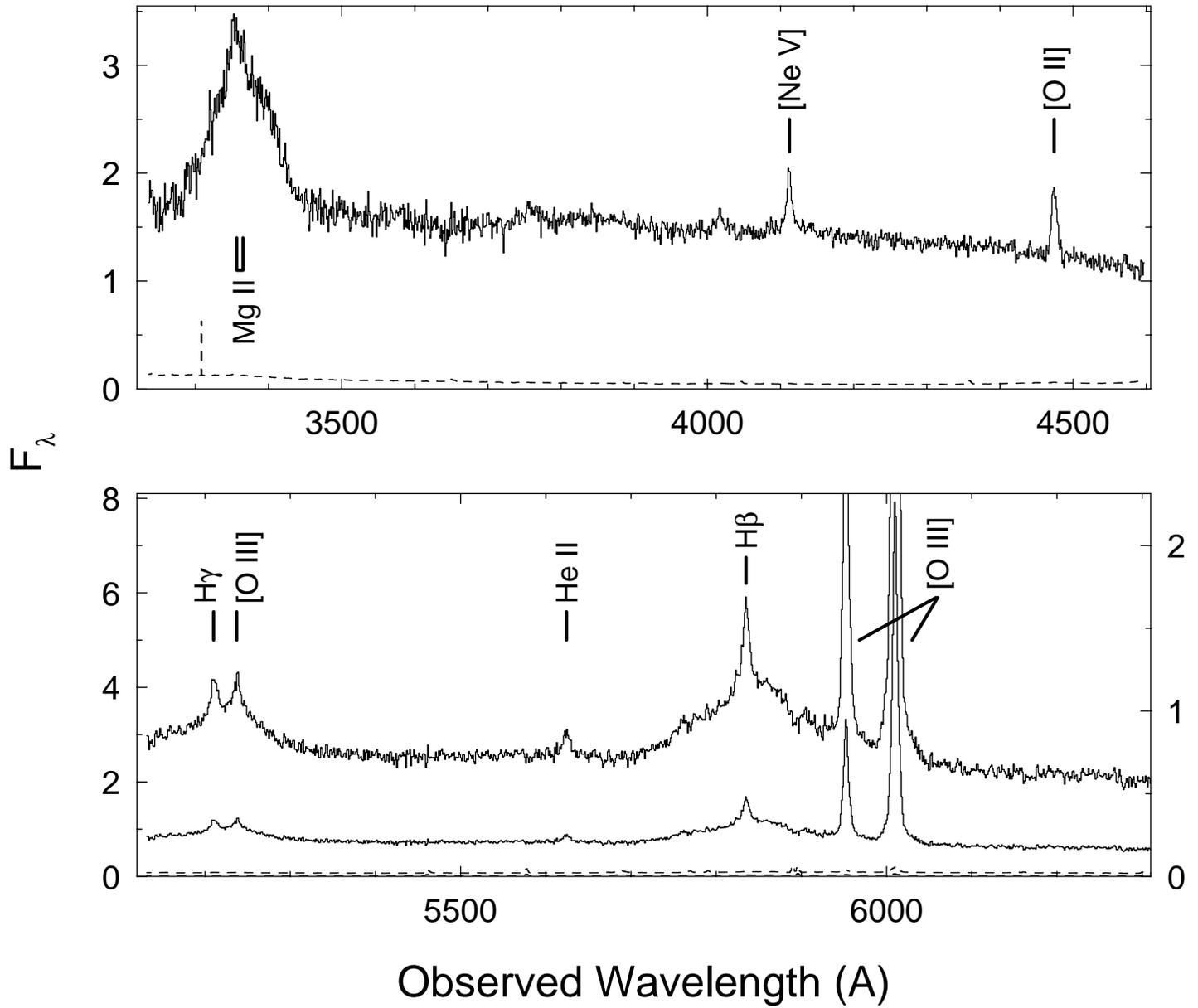
\end{document}